\documentclass[onecolumn]{aa}
\usepackage[varg]{txfonts}

\usepackage{graphicx}
\usepackage{natbib}
\usepackage{color}
\usepackage{subfig}
\usepackage[switch]{lineno}
\usepackage{xspace}
\usepackage{microtype}
\usepackage{amssymb}
\usepackage{mathtools}
\usepackage{multirow}
\def\link_col{blue}
\usepackage{xspace}
\usepackage{url}
\usepackage{wasysym}
\usepackage{caption}

\def\gray{$\gamma$-ray\xspace}
\def\grays{$\gamma$-rays\xspace}
\def\fermi{{\it Fermi}-LAT\xspace}

\begin{document}

\title{Tentative evidence of spatially extended GeV emission from SS433/W50}
\titlerunning{\fermi detection of SS433/W50}
\author{Xiao-Na Sun\inst{1,2}
\and Rui-Zhi Yang\inst{3}
\and Bing Liu\inst{1,3}
\and Shao-Qiang Xi\inst{1,2}
\and Xiang-Yu Wang\inst{1,2}
}
\institute{School of Astronomy and Space Science, Nanjing
University, Nanjing 210093, China\\
\and Key laboratory of Modern Astronomy and Astrophysics, Nanjing
University, Ministry of Education, Nanjing 210093, China\\
\and Max-Planck-Institut f\"ur Kernphysik, Saupfercheckweg 1, 69117 Heidelberg, Germany\\
}

\abstract {
We analyze 10 years of \fermi data towards the SS433/W50 region.
With the latest source catalog and diffuse background models, the \gray excess from SS433/W50 is detected with a significance of $\sim 6 \sigma$ in the photon energy range of 500 MeV - 10 GeV. Our analysis indicates that an extended flat disk morphology is preferred over a point-source description, suggesting that the GeV emission region is much larger than that of the TeV emission detected by HAWC. The size of the GeV emission is instead consistent with the extent of the radio nebula W50, a supernova remnant being distorted by the jets, so we suggest that the GeV emission may originate from this supernova remnant. The spectral result of the GeV emission is also consistent with an supernova remnant origin.
We also derive the GeV flux upper limits on the TeV emission region, which put moderate constrains on the leptonic models to explain the multiwavelength data.
}
 \maketitle

\section{Introduction}
\label{sec:intro}
SS 433 is a Galactic X-ray binary system containing a compact object (either a black hole or a neutron star) accreting matter from a companion star \citep{Gies02, Eikenberry01}.
{\bf It is the prototype of microquasars and has been studied intensively due to its exceptional brightness and power \citep[for a review, see e.g.,  ][]{Mirabel99}.}
It is located close to the geometric centre of the radio nebula W50, which is most likely a supernova remnant (SNR). Several X-ray hotspots located west of the central binary (w1 and w2) and east (e1, e2, e3) are observed \citep{Safi97}.
These regions are bright at X-rays due to the interaction between the jets and the ambient medium \citep{Safi97, Safi99, Brinkmann07}.
Two bright radio ears to the east and west of the nebula are generally believed to be punched out by two oppositely directed relativistic jets {\bf \citep{Abell79, Fabian79, Margon79, Milgrom79,Dubner98}} moving outwards from SS433 \citep{Margon84,Elston87, Goodalla, Goodall11, Farnes17}.
The entire radio morphology of the SS433/W50 system is thus seen as an approximately elliptical shell that is $2^\circ \times 1^\circ$ in projection on the sky \citep{Geldzahler80}.
S433/W50 lies at a distance of $5.5 \pm 0.2$ kpc \citep{Blundell04, Lockman07}, implying a physical size of $\rm 180\ pc  \times 80\ pc$ \citep{Goodalla}.

\citet{Bordas15} reports for the first time the GeV emission is tentatively associated with  SS433/W50  using five years of \fermi data.
Recently, the HAWC telescope detected very-high-energy (VHE) emission ($\rm \sim  20\ TeV$) around two X-ray hotspots e1 and w1 regions with a significance of $5.4\sigma$ \citep{zhou18}. The spatial extension analysis of the two hotspots yield upper limits on the angular size of the emission regions being $0.25^\circ$ for the east hotspot and  $0.35^\circ$ for the west hotspot
at 90\% confidence \citep{zhou18}. The spectral results indicate that the broadband spectral energy distribution (SED) at the e1 region can be naturally explained within a pure leptonic model, in which the same population of relativistic electrons with energies up to hundreds of TeV produce the radio to X-ray emission by the synchrotron radiation and up-scatter photons from the cosmic microwave background (CMB) to the TeV \grays.
\citet{Xing19} analyzes ten years of \fermi data using the preliminary LAT 8-year point source list (FL8Y)\footnote{https://fermi.gsfc.nasa.gov/ssc/data/access/lat/fl8y/} and finds excess GeV emission around the western X-ray hotspot w1 region.
However, we note that the point source FL8Y J1913.3+0515, which is consistent with the position of W50 (SNR G039.7-02.0), has been included in their background model, leading to a reduction of the excess \gray emission from the north-east region.
\citet{Rasul19} focuses on the temporal behavior of GeV emission and finds tentative evidence for periodicity at the precession period.
They analyze nine years of \fermi data using the \fermi 4-year catalog \citep[3FGL,][]{Acero15} and the foreground models {\it gll\_iem\_v06.fits} and {\it iso\_P8R3\_SOURCE\_V2.txt}\footnote{\url{https://fermi.gsfc.nasa.gov/ssc/data/access/lat/BackgroundModels.html}}.

Different from the previous works, in this paper, we use the most recent \fermi products to reanalyze the SS433/W50 region.
We focus on  the spatial extension of the GeV emission using multiple spatial templates. In Section~\ref{sec:spatial_analy} we perform a detailed spatial analysis on the \fermi data towards SS433. In Section~\ref{sec:spectral_analy}, we present the spectrum and discuss the radiation mechanism. Section~\ref{sec:conc} is the discussion and conclusion.

\section{Spatial analysis of the \fermi data}
\label{sec:spatial_analy}
We analyze \fermi Pass 8 database toward the SS433/W50 system from August 4, 2008 (MET 239557417) until December 18, 2018 (MET 566826221).
We select both the front and back converted photons at energies from $100~ {\rm MeV}$ to $10~ {\rm GeV}$.
A $14^ \circ \times 14^ \circ$ square region centred at the nominal position of SS433 is considered as the region of interest (ROI).
We use the "source" event class, recommended for individual source analysis, the recommended expression $\rm (DATA\_QUAL > 0) \&\& (LAT\_CONFIG == 1)$ to exclude time periods when some spacecraft event has affected the data quality.
To reduce the background contamination from the Earth's albedo, only the events with zenith angles less than $90^{\circ}$ are included for the analysis.
The data have been processed through the current Fermitools from conda distribution\footnote{https://github.com/fermi-lat/Fermitools-conda/} together with the latest version of the instrument response functions (IRFs) {\it P8R3\_SOURCE\_V2}.
We utilize the python module that implements a maximum likelihood optimization technique for a standard binned analysis\footnote{\url{https://fermi.gsfc.nasa.gov/ssc/data/analysis/scitools/python_tutorial.html}}.
For the complex structures at low latitude, it's more reasonable to re-optimize for {\bf Test Statistic (TS)} fits through setting the keyword tsmin = true.

In our background model, we use \fermi 8-year catalog \citep[4FGL,][]{Fermi19} by running the make4FGLxml script\footnote{\url{https://fermi.gsfc.nasa.gov/ssc/data/analysis/user/make4FGLxml.py}} within the region of ROI enlarged by $5^{\circ}$.
We leave the normalizations and spectral indices free for all sources within 6 degrees away from SS433.
For the foreground components, we use the latest Galactic diffuse model {\it gll\_iem\_v07.fits} and isotropic emission model {\it iso\_P8R3\_SOURCE\_V2\_v1.txt}\footnote{\url{https://fermi.gsfc.nasa.gov/ssc/data/access/lat/BackgroundModels.html}} with their normalization parameters free.

Generally, high energy maps with higher angular resolution are more suitable for the spatial analysis, but low statistics in the higher energy range may prevent any improvement.
Therefore, to balance the effects of these two factors, we select the data in the energy range of 500 MeV - 10 GeV for the analysis.
We use the {\it gttsmap} tool to evaluate a $2.5^{\circ}\times 2.5^{\circ}$ residual TS map by removing the contribution from all the known sources in our background model defined above. The TS value for each pixel is defined as ${\rm TS}=-2({\rm ln}L_0-{\rm ln}L_1)$, where $L_0$ is the maximum-likelihood value for null hypothesis and $L_1$ is the maximum likelihood with the additional source under consideration.
The spectral type of all the added sources for the likelihood ratio test in this section are assumed to be a simple power-law.
As shown in  \figurename~\ref{fig1}, a strong \gray excess near the SS433's nominal position is apparent after the fitting and subtraction of \grays from the background sources by performing the TS analysis.

At the eastern interaction region, there’s almost no GeV emission, which is roughly consistent with the results reported by \citet{Bordas15}, while TeV excess has been detected by HAWC’s observation in this region.
\citet{zhou18} has argued that the TeV emission originates from the jet termination shock, thus the above difference indicates that the GeV emission likely originates from the other regions, e.g., one of the possibilities, the SNR W50 itself.

\subsection{Single point-like source model}
We add a point-like source model encompassing the SS433's nominal position into our background model, and optimize the localization using the {\it gtfindsrc} tool. The derived best-fit position of the excess above 500 MeV is [R.A.$ = 288.02^{\circ}$, Dec.$ = 5.18^{\circ}$] (the "diamond" in \figurename~\ref{fig1}), with 1$\sigma$ and 2$\sigma$ error radii of 0.4$^{\circ}$ and 0.6$^{\circ}$, and $0.2^\circ$ away from the nominal position.
Moreover, the SS433's nominal position, e1 and w1 lie within 2$\sigma$ positional errors of the best-fit position.
As shown in \tablename~\ref{tab1}, under the assumption of single point-like source model, the significance of the excess \gray emission is $\rm TS = 26$ (${\bf\sim 5\sigma}$) and the best-fit photon index is $3.31 \pm 0.02$.

In the following, to investigate the morphology and extension of the \gray emission, we consider several spatial templates as the alternative hypothesis, and the above single point-like source model as the null hypothesis. Then we compare the overall maximum likelihood of the alternative hypothesis ($L$) with that of the null hypothesis ($L_{0}$), and define the significance of the alternative hypothesis model $-2({\rm ln}L_{0}-{\rm ln}L)$ following the paper \citet{Lande12}.

\subsection{Two point-like source model}
The e1 and w1 regions detected by HAWC \citep{zhou18} lie within 2$\sigma$ positional errors of the best-fit position, we cannot yet rule out the possibility that the \gray emission originates from the regions of the excess TeV emission.
To study whether the extended nature of this GeV emission is caused by a superposition of two separate point-like sources, we calculate maximum likelihood value for the most possible combination of two point-like sources, i.e., one point-like source at the e1 location and one point-like source at the w1 position (see \figurename~\ref{fig1}).
Given that $\rm \Delta TS = 3.4$ (${\bf < 2\sigma}$), this model is disfavored statistically.
Thus the GeV emission probably does not originate from the regions of the TeV emission.

\subsection{Uniform disk tests for spatial extension}
To investigate whether the \gray excess is extended beyond a point source spatially, we assume that the \gray source has uniform surface brightness inside the SNR boundary, and produce several uniform disk templates centered at the best-fit position with various radii from $0.1^\circ$ to $0.7^\circ$ in steps of $0.025^\circ$.
{\bf The derived significance of the excess \gray emission is $6 \sigma$ with 1 additional degrees of freedom relative to single point source.}
As shown in \figurename~\ref{fig2}, the extension likelihood is peaked at the disk radius of $\rm R_{disk} = 0.45^{\circ} \pm 0.06^{\circ}$ ($1\sigma$ error), which is roughly compatible with the extension of the radio emission \citep{Geldzahler80} longitudinally.
The peak significance of $-2({\rm ln}L_{0}-{\rm ln}L) = 12\ {\bf (\sim 3.5\sigma)}$ is a tentative evidence that the excess \gray emission at the SS433/W50 system is spatially extended.
The photon index is $2.62 \pm 0.02$, and \gray luminosity can be estimated as $\sim 6.48 \times 10^{34} (D/5.5 \ \rm kpc)^2 \ \rm erg/s$.

\subsection{Radio and X-ray templates}
To check if the \gray emission distribution traces the observed radio and X-ray emission, we consider three additional templates. We create two radio emission templates based on the measurements of the Green Bank Observatory at 4.8 GHz with and without two lateral "ears".
And the X-ray template is produced from the observation with ROSAT PSPC at 1.2 keV, taken from the Virtual Observatory $\it SkyView$\protect\footnote{\protect\url{https://skyview.gsfc.nasa.gov/current/cgi/titlepage.pl}}.
The contours of the radio and X-ray emission are overlaid in \figurename~\ref{fig1}, in white and magenta, respectively.
In \tablename~\ref{tab1}  we list the fitting results for the three templates using the binned likelihood analysis.
The significances of the two radio templates relative to single point source model, both with $\rm \Delta TS < 1$, shows no  improvement.
The significance derived from the case of the X-ray template is significantly lower than the single point source model.
This is expected since the excess \gray emission distribution is basically inconsistent with the region of the X-ray emission.
Therefore, we reject the hypothesis that the \gray emission originates from the extended region of the X-ray emission.

\subsection {Extension analysis using alternative diffuse background model}
\label{sec:alt_mol}
{We also use the previous Galactic interstellar emission model {\it gll\_iem\_v06.fits} \footnote{\url{https://fermi.gsfc.nasa.gov/ssc/data/access/lat/BackgroundModels.html}} to replace the {\it gll\_iem\_v07.fits} in our background model, and repeat the above spatial analyses.
In this case, the morphology of the TS map is roughly consistent with the one derived from the latest Galactic interstellar emission model.
Under the assumption of single point-like source model, the derived best-fit position of the excess above 500 MeV is [R.A.$ = 287.91^{\circ}$, Dec.$ = 5.13^{\circ}$], which is $0.15^\circ$ away from the nominal position, and the significance is ${\rm TS=35}\ {\bf (6\sigma)}$.
An uniform disk template ($\rm R_{disk} \sim 0.45^{\circ}$) can fit the \gray excess with an improvement of $\rm \Delta TS=24\ {\bf (\sim 5\sigma)}$ relative to the single point-like source model.
This shows that  inclusion of the latest diffuse background model {\it gll\_iem\_v07.fits} significantly reduces the detection significance, as well as the detected \gray flux. A dedicated analysis on the possible systematic errors due to the diffuse background models is needed.

\section{Spectral analysis}\label{sec:spectral_analy}
The results from spatial analysis show that the uniform circle disk template with a radius of $\sim 0.45^\circ$ is preferred over the point source for the \gray emission towards the SS433/W50 system, with an improvement of {\bf 3.5 to 5$\sigma$}.
{\bf Thus we fix the $0.45^\circ$ uniform circle disk as the spatial model of the extended \gray emission towards the SS433/W50 system to extract the spectral information.
The logparabola function, $\frac{dN}{dE} \propto \left(\frac{E}{E_{\rm b}}\right)^{-(\alpha+\beta~{\rm log}~ (E/E_{\rm b}))}$, can best fit the global spectral shape of the \gray emission with a $\rm TS=39\ {\bf (\sim 6.2\sigma)}$, and the parameters are $\alpha = 2.03 \pm 0.01$, $\beta =0.27 \pm 0.02 $, and $E_{\rm b} = 0.9 \pm 0.004\ \rm GeV$.}

\subsection{Modelling the spectral energy distribution}
To investigate the origin of the extended GeV emission and the underlying particle spectra that give rise to the observed spectrum of photons, we derive the spectral energy distribution (SED) of SS433/W50 via the maximum likelihood estimation in seven logarithmically spaced energy bins from 80 MeV - 10 GeV.
For the first and last energy bands, which are detected with a significance of less than $2\sigma$, we only free the normalization parameters in the iterative model derived from above spatial and spectral fits and calculate the upper limits within $3\sigma$ confidence level. The energy dispersion correction is applied to the first energy bin.
The derived SED of SS433/W50 in this work along with the SED from \citet{Bordas15} are plotted in \figurename~\ref{fig3}. We notice that while the spectral shape is consistent with that in \citet{Bordas15}, the normalization is reduced by a factor of more than 2. This is consistent with a lower detection significance derived in this work (about $5\sigma$ for point source for 10 years data in this work, in comparison with more than $7\sigma$ for 6 years data in  \citet{Bordas15}). We attribute the difference also to the different choice of the galactic diffuse background.

We use {\it Naima}\footnote{\url{http://naima.readthedocs.org/en/latest/index.html#}} \citep{naima} to fit the SED. Naima is a numerical package that includes a set of non-thermal radiative models and a spectral fitting procedure.
The code allows us to implement different functions and perform Markov chain Monte Carlo \citep[MCMC;][]{Foreman13} fitting of non-thermal radiative processes to the data.

To phenomenologically interpret the SED in the GeV band, we adopt both an exponential cut-off power-law (ECPL) and a more complex broken power-law (BPL) distribution of the radiating electrons or protons. The former is in the form of
\begin{equation}
N(E) = A~E^{-\alpha}~{\rm exp}\left[-\left(\frac{E}{E_{\rm cutoff}}\right)^{\rm \beta}\right],
\label{equ:ecpl}
\end{equation}
treating $A$, $\alpha$, $E_{\rm cutoff}$, and $\beta$ as free parameters for the fit, and the later is
\begin{equation}
N(E) =
\begin{cases}
AE^{-\alpha_{\rm 1}} & \quad  E\leq E_{\rm break}\\
A E_{\rm break}^{(\alpha_{\rm 2}-\alpha_{\rm 1})}~E^{-\alpha_{\rm 2}} & \quad  E>E_{\rm break},\\
\end{cases}
\label{equ:bpl}
\end{equation}
with the $A$, $\alpha_{1}$, $E_{\rm break}$, and $\alpha_{2}$ left as free parameters.

In the hadronic scenario, we attribute the observed \grays to the decay of neutral pions produced by the pp collision between the relativistic protons (and nuclei) in the SNR W50 and the protons in the interstellar medium, using the cross-section approximation of \citet{Kafexhiu14}.
The average number density of the target protons is assumed to be $\rm 4.0\ cm^{-3}$ as in \citet{Bordas15}.
On the other hand, we test the leptonic scenario assuming the \gray emission is produced by the IC up-scattering of seed photons, or through the bremsstrahlung of the relativistic electrons embedded in the SNR W50 or matter from the surrounding medium.
For the interstellar radiation field of the IC, except for the Cosmic Microwave Background (CMB), we also adopt the infrared and optical emissions from dust and starlight of the local interstellar radiation field calculated by \citet{Popescu17}.
For the relativistic bremsstrahlung, we assume the target particle density of $\rm 4\ cm^{-3}$ \citep{Bordas15}.
We adopt the formalism described in \citet{Aharonian81} and \citet{Khangulyan14} for the IC photon spectrum calculation and the relativistic bremsstrahlung spectra from \citet{Baring99}, respectively.

Table~\ref{tab2} lists the best-fit parameters as well as the total energy of the electrons or protons, $W_{\rm e/p}$, with $1 \sigma$ statistical errors, searched using the maximum log-likelihood (MML). Figure~\ref{fig3} shows the radiation models with the maximum likelihood values correspondingly.
It is evident that the $\pi^{0}$ decay model is the best one to match the flux points, which reveal a pion-bump feature. The parameters are $\alpha_{1} = 2.1 \pm 0.2$, $\alpha_{2} = 4.0 \pm 0.4$, $\rm E_{break} = 2.4^{+0.4}_{-0.3}\ GeV$, and the total energy of the high-energy protons $\rm W_{p} = (1.14 \pm 0.05) \times 10^{50}\ erg$. 
The leptonic model of GeV emission leads to a significant decrease in the maximum likelihood of the fitting, as shown in \tablename~\ref{tab2}.
The constraints on leptonic models are mainly from the first two energy bins, which can hardly be addressed by either IC or Bremsstrahlung process except we assume a very sharp low energy cutoff of the low energy electron spectrum. Such a hadronic origin can also explain the lower significance of the spatial analysis results by using radio templates. The radio emissions are mainly from the synchrotron radiation of electrons and can have a different spatial distribution than that of the proton content which is responsible for the \gray emission.

\subsection{Upper limits on  GeV emission from the e1 region}
Although no GeV emission is significantly detected at the east hotspot region e1 of SS433/W50, the upper limit flux  can provide  constrains on the leptonic model reported in \citet{zhou18}.
We add a point-like source  with a power-law spectral type at the e1 position into our iterative model, and obtain upper limits within a $2\sigma$ confidence level.
As shown in \figurename~\ref{fig4}, we consider the one zone leptonic model as in \citet{zhou18}, namely, the relativistic electrons scatter CMB photons to GeV - TeV energies via IC process (dot-dashed line) and produce radio to low energy \gray emissions by the synchrotron process (dashed line).
We also assume a power-law electron distribution with an exponential cutoff, that is, $dN/dE \propto E^{-\alpha}{\rm exp}(-E/E_{\rm max})$. The solid line presents the sum of the IC and synchrotron radiation.
We find the first upper limit in the energy range of 300 - 500 MeV can effectively constrain the $E_{\rm max}$, and the fitting results are roughly consistent with \citet{zhou18}.
The best-fit values of the parameters are $\alpha = 2.45^{+0.1}_{-0.2}$, $E_{\rm max} = 2.5^{+0.4}_{-0.3} \rm PeV$, $B = 15^{+3}_{-2} \rm \mu G$, and the total energy of the electrons with energy above 10 GeV is $(1.2 \pm 0.9)\times 10^{47} \rm erg$. It is interesting that the current best-fit model already predicts a turnover in the energy spectrum in the HAWC energy range. Such kind of feature can be directly tested with more HAWC observations in the future. And the next generation TeV \gray detectors, such as CTA \citep{Actis11} and LHAASO \citep{Sciascio16}, will definitely provide a clear test on the leptonic model in the termination region of the jet.

\section{Discussion and Conclusion}\label{sec:conc}
We have analyzed 10-year \fermi data towards SS433/W50 region and found tentative evidence for spatially extended GeV emission in this region (with a significance of 3.5 to $5\sigma$, depending on the choice of the diffuse background). The extension of the sub-GeV to GeV \gray emission are roughly consistent with the size of SNR W50. The SED of these \gray emission reveals a pion-bump shape and can be well explained by the pion-decay emissions from a population of protons with a low break energy in the spectrum interacting with the ambient gas. Such a slightly accelerated proton population can be  injected along the jet or at the jet termination shock \citep{Bordas15}. The acceleration mechanism should be  rather inefficient, given that the \gray spectrum extends only up to a few GeVs. Recently, \citet{Rasul19} has found tentative evidence for periodicity at the precession period of the GeV emissions in SS433. If this is true, at least part of these GeV emissions should be produced in the basement of the jet, rather than in the termination region or the SNR. However, we note that the significance of the variability is only about $3\sigma$. And we note that their analysis are performed with old diffuse background models. As mentioned in Section~\ref{sec:alt_mol}, the update to the latest diffuse background models has significantly reduced the detection significance. A further detailed systematic investigation of diffuse background models in this region are needed to clarify these features.  On the other hand, due to the spatial coincidence of the sub-GeV to GeV  \gray emission and the SNR itself, it is more likely that these protons are accelerated in the SNR. We also found that a broken power-law proton spectrum can best fit the \gray SED. Such a broken power-law shape spectrum is also observed in other mid-age SNRs, such as IC433 and W44 \citep{Ackermann13}. However the break energy here is much lower than those in other mid-age SNRs. Such a low-energy break may reflects the inefficient acceleration in this SNR due to the presence of the jet. On the other hand, \citet{Malkov11} has explained the energy break in these mid-age SNRs as the result of Alfven waves evanescence caused by strong ion-neutral collisions. In this case the break energy is determined by the neutral and ion density, as well as the gas temperatures, which can be very different in W50 due to the influence of the jet.

Due to the different spatial morphology, the \gray emissions observed by \fermi may have a different origin than that of the TeV emissions observed by HAWC which is believed to originate from the termination region of the microquasar jet \citep{zhou18}. Under this assumption we also extract the upper limits on the TeV emission region in \fermi energy band. We found that the current data can already put interesting constraints on the maximum electron energy in the leptonic model. It is interesting that the current best-fit leptonic model predicts an turnover in the energy spectrum in HAWC energy range, which can be tested by the further data release of HAWC, as well as the observations of the forthcoming CTA and LHASSO.

\begin{figure}[ht]
\centering
\includegraphics[scale=0.4]{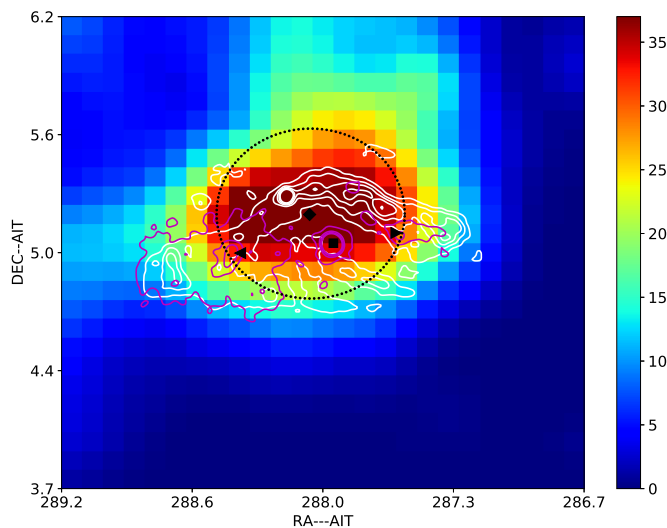}
\caption {$2.5^{\circ} \times 2.5^{\circ}$ TS residual map near the SS433/W50 system above 500 MeV, with pixel size corresponding to $0.1^{\circ} \times 0.1^{\circ}$.
The square shows the nominal position of SS433, and the diamond indicates the best-fit position.
The left and right triangles (e1 \& w1) mark the fixed positions for the fitting of the VHE excesses detected by HAWC.
The contours in white to show the main features of the studied region using the radio observation, smoothed with a Gaussian kernel of $0.1^ \circ$. Contours start at 10 mJy beam$^{-1}$ and increase in steps of 40 mJy beam$^{-1}$.
The magenta contours show the X-ray emission, smoothed with a Gaussian filter of $0.17^ \circ$.
The dashed circle with a radius of $0.45^{\circ}$ shows the size of the assumed uniform disk used for spatial analysis.
}
\label{fig1}
\end{figure}

\begin{figure}[ht]
\centering
\includegraphics[scale=0.4]{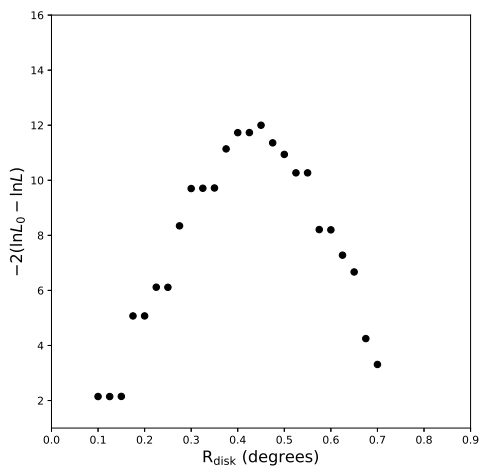}
\caption {
The significance of the uniform disk models with various radii relative to the single-point source model. The maximum likelihood radius of the disk is at $\rm R_{disk} \sim 0.45^{\circ}$.
}
\label{fig2}
\end{figure}

\begin{figure}[ht]
\centering
\includegraphics[scale=0.4]{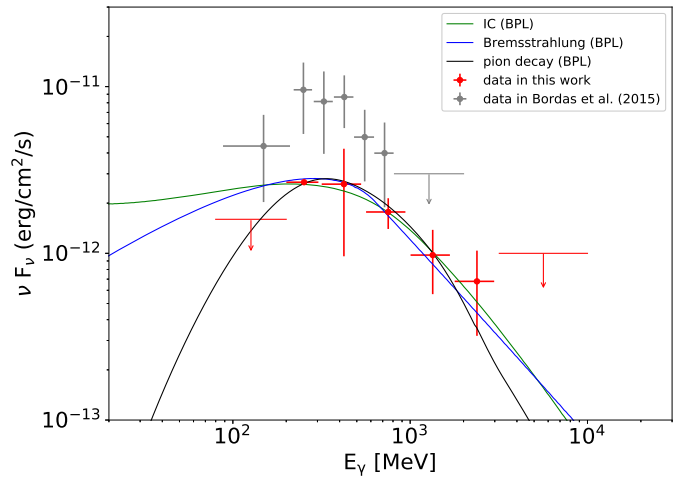}
\caption {SED of the GeV emission component of the SS433/W50 system for the $0.45^{\circ}$ uniform circle disk spatial model. The upper limits are calculated within a $3\sigma$ confidence level. The grey data are taken from the Figure $3$ in \citet{Bordas15}.}
\label{fig3}
\end{figure}

\begin{figure}[ht]
\centering
\includegraphics[scale=0.2]{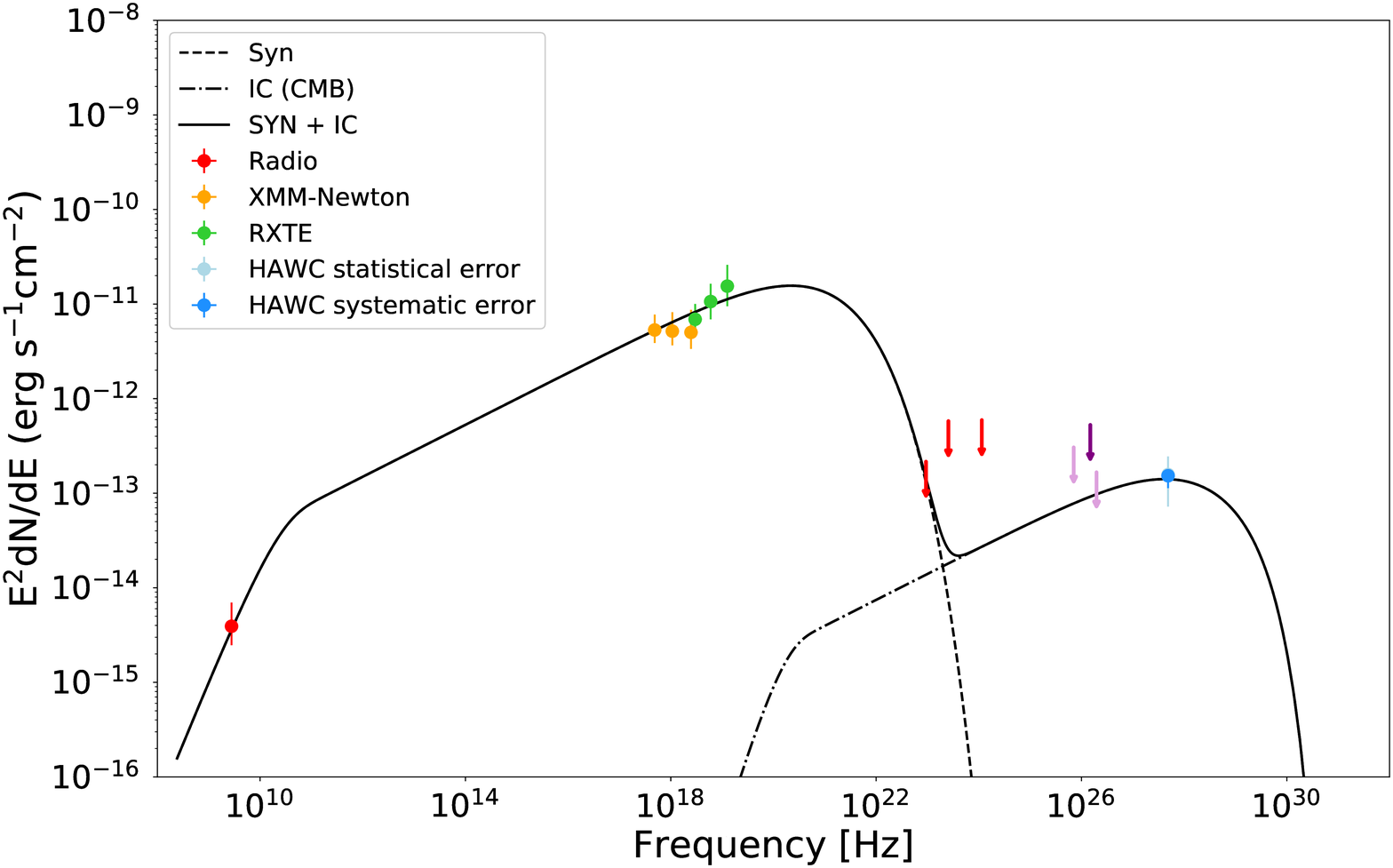}
\caption {SED of the eastern emission region e1. The red upper limits are derived in this work within $2\sigma$ confidence level. Other data are taken from \citet{zhou18} directly. Radio to low energy \gray photons are fitted with a synchrotron emission in a magnetic field (dashed line), and the GeV to TeV emission through IC scattering of the CMB (dot-dashed line).
The solid line represents the sum of the synchrotron and IC emission.
}
\label{fig4}
\end{figure}

\begin{table*}
\centering
\caption{Spatial analysis results at $500 \rm\ MeV-10\rm\ GeV$}
\begin{tabular}{lcccccr}
\hline
\hline
Spatial Model & Energy Flux&  Power-law Index & $-2({\rm ln}L_{0}-{\rm ln}L)$\tablefootmark{b}&DoF\tablefootmark{c}\\
 &($\times 10^{-11}{\rm \ erg\ cm^{-2}\ s^{-1}}$) & & &\\
 \hline
A\tablefootmark{a} & 2.80 $\pm$ 0.13 & 3.31 $\pm$ 0.02 & 0& 0\\ 
\\
\multirow{2}*{$\rm e1+w1$}& 0.65 $\pm$ 0.17 & 2.88 $\pm$ 0.10 & \multirow{2}*{3.4} &\multirow{2}*{2}\\
& 2.35 $\pm$ 0.83 & 3.29 $\pm$ 0.24 &  &\\
\\
Uniform circle disk ($0.45^\circ$) & 1.79 $\pm$ 0.05 & 2.62 $\pm$ 0.02 & 12&1\\
\\
Radio (4.8 GHz) & 2.94 $\pm$ 0.16 & 3.23 $\pm$ 0.04 & 0.1&0\\
\\
Radio (4.8 GHz, no ears) & 0.87 $\pm$ 0.01 & 2.45 $\pm$ 0.03 & < 0&0\\
\\
X-ray (240 PHz) & $\sim 0$ & 2.03 $\pm$ 0.69 & < 0&0\\
\hline
\hline
\end{tabular}
\tablefoot{
\tablefoottext{a}{Model A corresponds to point source model with a power-law spectrum at the best-fit location in \figurename~\ref{fig1}.}
\tablefoottext{b} $-2({\rm ln}L_{0}-{\rm ln}L)$ represents the significance of the alternative hypothesis relative to the model A (null hypothesis).
\tablefoottext{c}{Additional degrees of freedom.}\\
Each uniform disk model is centered at the best-fit position.
All the associated uncertainty refers to the 68\% statistic errors.
Details see the context in Section~\ref{sec:spatial_analy}.
}
\label{tab1}
\end{table*}

\begin{table*}
\caption{SED fit results for different radiation models.}
\centering
\begin{tabular}{ccccccc}
\hline\hline
Radiation&Parent particle& $W_{\rm e/p}$\tablefootmark{a} &$\rm \alpha\ or\ \alpha_{1}$ &$\rm E_{\rm cutoff}\ or\ E_{\rm break}$ &$\rm \beta\ or\ \alpha_{2}$&$\rm MLL$\tablefootmark{b}   \\
model &distribution&($\times 10^{49}$ erg) & &($\rm GeV$) & \\ [0.1cm]
\hline
\multirow{3}*{pp}& ECPL &11.9$^{+0.4}_{-0.3}$&2.1$\pm$0.2&2.7$^{+0.4}_{-0.3}$&1 (fixed)&-1.6  \\[0.1cm]
&ECPL &12.1$^{+1.1}_{-0.8}$&2.1$^{+0.3}_{-0.2}$&2.8$^{+0.6}_{-0.4}$&1.1$\pm$0.3&-1.6  \\[0.1cm]
&BPL &11.4$\pm$0.5&2.1$\pm$0.2&2.4$^{+0.4}_{-0.3}$&4.0$\pm$0.4&-1.0  \\[0.1cm]
\hline
\multirow{3}*{IC}& ECPL &2.8$^{+0.3}_{-0.2}$&1.5$\pm$0.1&9.5$\pm$1.2&1 (fixed)&-31.4 \\[0.1cm]
&ECPL &3.7$\pm$0.3&1.7$\pm$0.1&12$^{+1.8}_{-1.2}$&5.3$\pm$0.7&-13.3  \\[0.1cm]
&BPL &3.4$^{+0.3}_{-0.5}$&1.6$\pm$0.1&8.3$^{+1.4}_{-0.8}$&6.4$^{+1.4}_{-0.8}$&-10.3  \\[0.1cm]
\hline
\multirow{3}*{Rel. bremsstrahlung}& ECPL &1.6$\pm$0.1&1.6$\pm$0.2&1.0$^{+0.4}_{-0.2}$&1 (fixed)&-10.6  \\[0.1cm]
&ECPL &1.5$\pm$0.1&1.7$\pm$0.2&1.4$\pm$0.3&1.8$^{+0.9}_{-0.5}$&-10.8  \\[0.1cm]
&BPL &1.5$\pm$0.1&1.5$\pm$0.3&0.7$^{+0.3}_{-0.1}$&3.7$^{+0.7}_{-0.4}$&-7.5  \\[0.1cm]

\hline
\end{tabular}
\tablefoot{All errors  correspond to a $1\sigma$ confidence level.
\tablefoottext{a}{Total energy of the electrons or protons. The $W_{\rm p}$ is calculated based on the protons $\rm > 1\ GeV$ and the $W_{\rm e}$ is calculated based on the electrons $\rm > 100\ MeV$.}
\tablefoottext{b}{Maximum log-likelihood {\bf (MLL)}.}
}
\label{tab2}
\end{table*}

\section*{Acknowledgements}
X.N. Sun thanks Yunfeng Liang for helpful discussions.
This work is supported by the NSFC under grants 11625312 and 11851304, and the National Key R\&D program of China under the grant 2018YFA0404203.

\bibliographystyle{aa}
\bibliography{ms}

\begin{thebibliography}{37}
\expandafter\ifx\csname natexlab\endcsname\relax\def\natexlab#1{#1}\fi

\bibitem[{{Abell} \& {Margon}(1979)}]{Abell79}
{Abell}, G.~O. \& {Margon}, B. 1979, \nat, 279, 701

\bibitem[{{Abeysekara} {et~al.}(2018){Abeysekara}, {Albert}, {Alfaro},
  {Alvarez}, {{\'A}lvarez}, {Arceo}, {Arteaga-Vel{\'a}zquez}, {Avila Rojas},
  {Ayala Solares}, {Belmont-Moreno}, {BenZvi}, {Brisbois}, {Caballero-Mora},
  {Capistr{\'a}n}, {Carrami{\~n}ana}, {Casanova}, {Castillo}, {Cotti},
  {Cotzomi}, {Couti{\~n}o de Le{\'o}n}, {De Le{\'o}n}, {De la Fuente},
  {D{\'{\i}}az-V{\'e}lez}, {Dichiara}, {Dingus}, {DuVernois}, {Ellsworth},
  {Engel}, {Espinoza}, {Fang}, {Fleischhack}, {Fraija}, {Galv{\'a}n-G{\'a}mez},
  {Garc{\'{\i}}a-Gonz{\'a}lez}, {Garfias}, {Gonz{\'a}lez-Mu{\~n}oz},
  {Gonz{\'a}lez}, {Goodman}, {Hampel-Arias}, {Harding}, {Hernandez}, {Hinton},
  {Hona}, {Hueyotl-Zahuantitla}, {Hui}, {H{\"u}ntemeyer}, {Iriarte},
  {Jardin-Blicq}, {Joshi}, {Kaufmann}, {Kar}, {Kunde}, {Lauer}, {Lee},
  {Le{\'o}n Vargas}, {Li}, {Linnemann}, {Longinotti}, {Luis-Raya},
  {L{\'o}pez-Coto}, {Malone}, {Marinelli}, {Martinez}, {Martinez-Castellanos},
  {Mart{\'{\i}}nez-Castro}, {Matthews}, {Miranda-Romagnoli}, {Moreno},
  {Mostaf{\'a}}, {Nayerhoda}, {Nellen}, {Newbold}, {Nisa}, {Noriega-Papaqui},
  {Pretz}, {P{\'e}rez-P{\'e}rez}, {Ren}, {Rho}, {Rivi{\`e}re},
  {Rosa-Gonz{\'a}lez}, {Rosenberg}, {Ruiz-Velasco}, {Salesa Greus}, {Sandoval},
  {Schneider}, {Schoorlemmer}, {Seglar Arroyo}, {Sinnis}, {Smith}, {Springer},
  {Surajbali}, {Taboada}, {Tibolla}, {Tollefson}, {Torres}, {Vianello},
  {Villase{\~n}or}, {Weisgarber}, {Werner}, {Westerhoff}, {Wood}, {Yapici},
  {Yodh}, {Zepeda}, {Zhang}, \& {Zhou}}]{zhou18}
{Abeysekara}, A.~U., {Albert}, A., {Alfaro}, R., {et~al.} 2018, \nat, 562, 82

\bibitem[{{Acero} {et~al.}(2015){Acero}, {Ackermann}, {Ajello}, {Albert},
  {Atwood}, {Axelsson}, {Baldini}, {Ballet}, {Barbiellini}, {Bastieri},
  {Belfiore}, {Bellazzini}, {Bissaldi}, {Blandford}, {Bloom}, {Bogart},
  {Bonino}, {Bottacini}, {Bregeon}, {Britto}, {Bruel}, {Buehler}, {Burnett},
  {Buson}, {Caliandro}, {Cameron}, {Caputo}, {Caragiulo}, {Caraveo},
  {Casandjian}, {Cavazzuti}, {Charles}, {Chaves}, {Chekhtman}, {Cheung},
  {Chiang}, {Chiaro}, {Ciprini}, {Claus}, {Cohen-Tanugi}, {Cominsky}, {Conrad},
  {Cutini}, {D'Ammando}, {de Angelis}, {DeKlotz}, {de Palma}, {Desiante},
  {Digel}, {Di Venere}, {Drell}, {Dubois}, {Dumora}, {Favuzzi}, {Fegan},
  {Ferrara}, {Finke}, {Franckowiak}, {Fukazawa}, {Funk}, {Fusco}, {Gargano},
  {Gasparrini}, {Giebels}, {Giglietto}, {Giommi}, {Giordano}, {Giroletti},
  {Glanzman}, {Godfrey}, {Grenier}, {Grondin}, {Grove}, {Guillemot}, {Guiriec},
  {Hadasch}, {Harding}, {Hays}, {Hewitt}, {Hill}, {Horan}, {Iafrate}, {Jogler},
  {J{\'o}hannesson}, {Johnson}, {Johnson}, {Johnson}, {Johnson}, {Kamae},
  {Kataoka}, {Katsuta}, {Kuss}, {La Mura}, {Landriu}, {Larsson}, {Latronico},
  {Lemoine-Goumard}, {Li}, {Li}, {Longo}, {Loparco}, {Lott}, {Lovellette},
  {Lubrano}, {Madejski}, {Massaro}, {Mayer}, {Mazziotta}, {McEnery},
  {Michelson}, {Mirabal}, {Mizuno}, {Moiseev}, {Mongelli}, {Monzani},
  {Morselli}, {Moskalenko}, {Murgia}, {Nuss}, {Ohno}, {Ohsugi}, {Omodei},
  {Orienti}, {Orlando}, {Ormes}, {Paneque}, {Panetta}, {Perkins},
  {Pesce-Rollins}, {Piron}, {Pivato}, {Porter}, {Racusin}, {Rando}, {Razzano},
  {Razzaque}, {Reimer}, {Reimer}, {Reposeur}, {Rochester}, {Romani},
  {Salvetti}, {S{\'a}nchez-Conde}, {Saz Parkinson}, {Schulz}, {Siskind},
  {Smith}, {Spada}, {Spandre}, {Spinelli}, {Stephens}, {Strong}, {Suson},
  {Takahashi}, {Takahashi}, {Tanaka}, {Thayer}, {Thayer}, {Thompson},
  {Tibaldo}, {Tibolla}, {Torres}, {Torresi}, {Tosti}, {Troja}, {Van Klaveren},
  {Vianello}, {Winer}, {Wood}, {Wood}, {Zimmer}, \& {Fermi-LAT
  Collaboration}}]{Acero15}
{Acero}, F., {Ackermann}, M., {Ajello}, M., {et~al.} 2015, \apjs, 218, 23

\bibitem[{{Ackermann} {et~al.}(2013){Ackermann}, {Ajello}, {Allafort},
  {Baldini}, {Ballet}, {Barbiellini}, {Baring}, {Bastieri}, {Bechtol},
  {Bellazzini}, {Blandford}, {Bloom}, {Bonamente}, {Borgland}, {Bottacini},
  {Brandt}, {Bregeon}, {Brigida}, {Bruel}, {Buehler}, {Busetto}, {Buson},
  {Caliandro}, {Cameron}, {Caraveo}, {Casandjian}, {Cecchi}, {{\c C}elik},
  {Charles}, {Chaty}, {Chaves}, {Chekhtman}, {Cheung}, {Chiang}, {Chiaro},
  {Cillis}, {Ciprini}, {Claus}, {Cohen-Tanugi}, {Cominsky}, {Conrad}, {Corbel},
  {Cutini}, {D'Ammando}, {de Angelis}, {de Palma}, {Dermer}, {do Couto e
  Silva}, {Drell}, {Drlica-Wagner}, {Falletti}, {Favuzzi}, {Ferrara},
  {Franckowiak}, {Fukazawa}, {Funk}, {Fusco}, {Gargano}, {Germani},
  {Giglietto}, {Giommi}, {Giordano}, {Giroletti}, {Glanzman}, {Godfrey},
  {Grenier}, {Grondin}, {Grove}, {Guiriec}, {Hadasch}, {Hanabata}, {Harding},
  {Hayashida}, {Hayashi}, {Hays}, {Hewitt}, {Hill}, {Hughes}, {Jackson},
  {Jogler}, {J{\'o}hannesson}, {Johnson}, {Kamae}, {Kataoka}, {Katsuta},
  {Kn{\"o}dlseder}, {Kuss}, {Lande}, {Larsson}, {Latronico}, {Lemoine-Goumard},
  {Longo}, {Loparco}, {Lovellette}, {Lubrano}, {Madejski}, {Massaro}, {Mayer},
  {Mazziotta}, {McEnery}, {Mehault}, {Michelson}, {Mignani}, {Mitthumsiri},
  {Mizuno}, {Moiseev}, {Monzani}, {Morselli}, {Moskalenko}, {Murgia},
  {Nakamori}, {Nemmen}, {Nuss}, {Ohno}, {Ohsugi}, {Omodei}, {Orienti},
  {Orlando}, {Ormes}, {Paneque}, {Perkins}, {Pesce-Rollins}, {Piron}, {Pivato},
  {Rain{\`o}}, {Rando}, {Razzano}, {Razzaque}, {Reimer}, {Reimer}, {Ritz},
  {Romoli}, {S{\'a}nchez-Conde}, {Schulz}, {Sgr{\`o}}, {Simeon}, {Siskind},
  {Smith}, {Spandre}, {Spinelli}, {Stecker}, {Strong}, {Suson}, {Tajima},
  {Takahashi}, {Takahashi}, {Tanaka}, {Thayer}, {Thayer}, {Thompson},
  {Thorsett}, {Tibaldo}, {Tibolla}, {Tinivella}, {Troja}, {Uchiyama}, {Usher},
  {Vandenbroucke}, {Vasileiou}, {Vianello}, {Vitale}, {Waite}, {Werner},
  {Winer}, {Wood}, {Wood}, {Yamazaki}, {Yang}, \& {Zimmer}}]{Ackermann13}
{Ackermann}, M., {Ajello}, M., {Allafort}, A., {et~al.} 2013, Science, 339, 807

\bibitem[{{Actis} {et~al.}(2011){Actis}, {Agnetta}, {Aharonian},
  {Akhperjanian}, {Aleksi{\'c}}, {Aliu}, {Allan}, {Allekotte}, {Antico},
  {Antonelli}, \& et~al.}]{Actis11}
{Actis}, M., {Agnetta}, G., {Aharonian}, F., {et~al.} 2011, Experimental
  Astronomy, 32, 193

\bibitem[{{Aharonian} \& {Atoyan}(1981)}]{Aharonian81}
{Aharonian}, F.~A. \& {Atoyan}, A.~M. 1981, \apss, 79, 321

\bibitem[{{Baring} {et~al.}(1999){Baring}, {Ellison}, {Reynolds}, {Grenier}, \&
  {Goret}}]{Baring99}
{Baring}, M.~G., {Ellison}, D.~C., {Reynolds}, S.~P., {Grenier}, I.~A., \&
  {Goret}, P. 1999, \apj, 513, 311

\bibitem[{{Blundell} \& {Bowler}(2004)}]{Blundell04}
{Blundell}, K.~M. \& {Bowler}, M.~G. 2004, \apjl, 616, L159

\bibitem[{{Bordas} {et~al.}(2015){Bordas}, {Yang}, {Kafexhiu}, \&
  {Aharonian}}]{Bordas15}
{Bordas}, P., {Yang}, R., {Kafexhiu}, E., \& {Aharonian}, F. 2015, \apjl, 807,
  L8

\bibitem[{{Brinkmann} {et~al.}(2007){Brinkmann}, {Pratt}, {Rohr}, {Kawai}, \&
  {Burwitz}}]{Brinkmann07}
{Brinkmann}, W., {Pratt}, G.~W., {Rohr}, S., {Kawai}, N., \& {Burwitz}, V.
  2007, \aap, 463, 611

\bibitem[{{Di Sciascio} \& {LHAASO Collaboration}(2016)}]{Sciascio16}
{Di Sciascio}, G. \& {LHAASO Collaboration}. 2016, Nuclear and Particle Physics
  Proceedings, 279, 166

\bibitem[{{Dubner} {et~al.}(1998){Dubner}, {Holdaway}, {Goss}, \&
  {Mirabel}}]{Dubner98}
{Dubner}, G.~M., {Holdaway}, M., {Goss}, W.~M., \& {Mirabel}, I.~F. 1998, \aj,
  116, 1842

\bibitem[{{Eikenberry} {et~al.}(2001){Eikenberry}, {Cameron}, {Fierce}, {Kull},
  {Dror}, {Houck}, \& {Margon}}]{Eikenberry01}
{Eikenberry}, S.~S., {Cameron}, P.~B., {Fierce}, B.~W., {et~al.} 2001, \apj,
  561, 1027

\bibitem[{{Elston} \& {Baum}(1987)}]{Elston87}
{Elston}, R. \& {Baum}, S. 1987, \aj, 94, 1633

\bibitem[{{Fabian} \& {Rees}(1979)}]{Fabian79}
{Fabian}, A.~C. \& {Rees}, M.~J. 1979, \mnras, 187, 13P

\bibitem[{{Farnes} {et~al.}(2017){Farnes}, {Gaensler}, {Purcell}, {Sun},
  {Haverkorn}, {Lenc}, {O'Sullivan}, \& {Akahori}}]{Farnes17}
{Farnes}, J.~S., {Gaensler}, B.~M., {Purcell}, C., {et~al.} 2017, \mnras, 467,
  4777

\bibitem[{{Foreman-Mackey} {et~al.}(2013){Foreman-Mackey}, {Hogg}, {Lang}, \&
  {Goodman}}]{Foreman13}
{Foreman-Mackey}, D., {Hogg}, D.~W., {Lang}, D., \& {Goodman}, J. 2013, \pasp,
  125, 306

\bibitem[{{Geldzahler} {et~al.}(1980){Geldzahler}, {Pauls}, \&
  {Salter}}]{Geldzahler80}
{Geldzahler}, B.~J., {Pauls}, T., \& {Salter}, C.~J. 1980, \aap, 84, 237

\bibitem[{{Gies} {et~al.}(2002){Gies}, {Huang}, \& {McSwain}}]{Gies02}
{Gies}, D.~R., {Huang}, W., \& {McSwain}, M.~V. 2002, \apjl, 578, L67

\bibitem[{{Goodall} {et~al.}(2011{\natexlab{a}}){Goodall}, {Alouani-Bibi}, \&
  {Blundell}}]{Goodall11}
{Goodall}, P.~T., {Alouani-Bibi}, F., \& {Blundell}, K.~M. 2011{\natexlab{a}},
  \mnras, 414, 2838

\bibitem[{{Goodall} {et~al.}(2011{\natexlab{b}}){Goodall}, {Blundell}, \& {Bell
  Burnell}}]{Goodalla}
{Goodall}, P.~T., {Blundell}, K.~M., \& {Bell Burnell}, S.~J.
  2011{\natexlab{b}}, \mnras, 414, 2828

\bibitem[{{Kafexhiu} {et~al.}(2014){Kafexhiu}, {Aharonian}, {Taylor}, \&
  {Vila}}]{Kafexhiu14}
{Kafexhiu}, E., {Aharonian}, F., {Taylor}, A.~M., \& {Vila}, G.~S. 2014, \prd,
  90, 123014

\bibitem[{{Khangulyan} {et~al.}(2014){Khangulyan}, {Aharonian}, \&
  {Kelner}}]{Khangulyan14}
{Khangulyan}, D., {Aharonian}, F.~A., \& {Kelner}, S.~R. 2014, \apj, 783, 100

\bibitem[{{Lande} {et~al.}(2012){Lande}, {Ackermann}, {Allafort}, {Ballet},
  {Bechtol}, {Burnett}, {Cohen-Tanugi}, {Drlica-Wagner}, {Funk}, {Giordano},
  {Grondin}, {Kerr}, \& {Lemoine-Goumard}}]{Lande12}
{Lande}, J., {Ackermann}, M., {Allafort}, A., {et~al.} 2012, \apj, 756, 5

\bibitem[{{Lockman} {et~al.}(2007){Lockman}, {Blundell}, \& {Goss}}]{Lockman07}
{Lockman}, F.~J., {Blundell}, K.~M., \& {Goss}, W.~M. 2007, \mnras, 381, 881

\bibitem[{{Malkov} {et~al.}(2011){Malkov}, {Diamond}, \& {Sagdeev}}]{Malkov11}
{Malkov}, M.~A., {Diamond}, P.~H., \& {Sagdeev}, R.~Z. 2011, Nature
  Communications, 2, 194

\bibitem[{{Margon}(1984)}]{Margon84}
{Margon}, B. 1984, \araa, 22, 507

\bibitem[{{Margon} {et~al.}(1979){Margon}, {Ford}, {Grandi}, \&
  {Stone}}]{Margon79}
{Margon}, B., {Ford}, H.~C., {Grandi}, S.~A., \& {Stone}, R.~P.~S. 1979, \apjl,
  233, L63

\bibitem[{{Milgrom}(1979)}]{Milgrom79}
{Milgrom}, M. 1979, \aap, 76, L3

\bibitem[{{Mirabel} \& {Rodr{\'{\i}}guez}(1999)}]{Mirabel99}
{Mirabel}, I.~F. \& {Rodr{\'{\i}}guez}, L.~F. 1999, \araa, 37, 409

\bibitem[{{Popescu} {et~al.}(2017){Popescu}, {Yang}, {Tuffs}, {Natale},
  {Rushton}, \& {Aharonian}}]{Popescu17}
{Popescu}, C.~C., {Yang}, R., {Tuffs}, R.~J., {et~al.} 2017, \mnras, 470, 2539

\bibitem[{{Rasul} {et~al.}(2019){Rasul}, {Chadwick}, {Graham}, \&
  {Brown}}]{Rasul19}
{Rasul}, K., {Chadwick}, P.~M., {Graham}, J.~A., \& {Brown}, A.~M. 2019, \mnras

\bibitem[{{Safi-Harb} \& {{\"O}gelman}(1997)}]{Safi97}
{Safi-Harb}, S. \& {{\"O}gelman}, H. 1997, \apj, 483, 868

\bibitem[{{Safi-Harb} \& {Petre}(1999)}]{Safi99}
{Safi-Harb}, S. \& {Petre}, R. 1999, \apj, 512, 784

\bibitem[{{The Fermi-LAT collaboration}(2019)}]{Fermi19}
{The Fermi-LAT collaboration}. 2019, arXiv e-prints
  {1902.10045}

\bibitem[{{Xing} {et~al.}(2019){Xing}, {Wang}, {Zhang}, {Chen}, \&
  {Jithesh}}]{Xing19}
{Xing}, Y., {Wang}, Z., {Zhang}, X., {Chen}, Y., \& {Jithesh}, V. 2019, \apj,
  872, 25

\bibitem[{{Zabalza}(2015)}]{naima}
{Zabalza}, V. 2015, Proc.~of International Cosmic Ray Conference 2015, in press

\end{thebibliography}

\end{document}